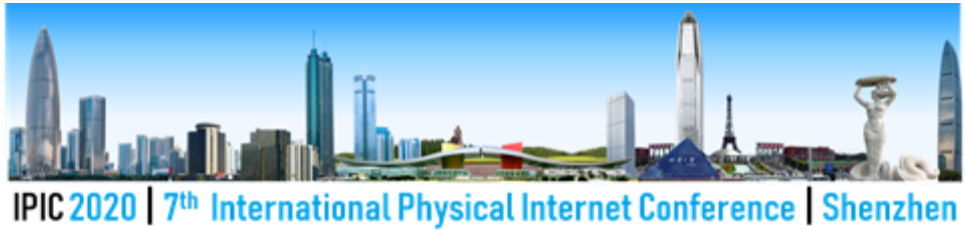

# Design and Evaluation of Routing Artifacts as a Part of the Physical Internet Framework

Steffen Kaup[1,4], André Ludwig[2], Bogdan Franczyk[1,3]

1. Leipzig University, Information Systems Institute, Leipzig, Germany
2. Kühne Logistics University, Computer Science in Logistics, Hamburg, Germany
3. Wrocław University of Economics, Wrocław, Poland
4. Mercedes-Benz AG, Group Research and Sustainability, Böblingen, Germany

*steffen.kaup@uni-leipzig.de; bogdan.franczyk@wifa.uni-leipzig.de; andre.ludwig@the-klu.org*

*Abstract*

"*Global freight demand will triple between 2015 and 2050, based on the current demand pathway*", as predicted in the Transport Outlook 2019 (Forum and International Transport Forum, 2019, p. 36). Based on the current traffic situation in the existing transport infrastructure, an increase in traffic on this scale is hardly conceivable. Hence, a revolutionary change in transport efficiency is urgently needed. One approach to tackle this change is to transfer the successful model of the Digital Internet for data exchange to the physical transport of goods: The so-called Physical Internet (PI, or π). The potential of the Physical Internet lies in dynamic routing, which increases the utilization of transport modalities, like trucks and vans, and makes transport more efficient. The main physical entities in the Physical Internet include π-nodes, π-containers and π-transporters. Previous concept transfers have identified and determined the π-nodes as routing entities. Here, the problem is that the π-nodes have no information about real-time data on transport vacancies. This leads to a great challenge for the π-nodes with regard to routing, in particular in determining the next best appropriate node for onward transport of the freight package. In the near future, it can be assumed that series production vehicles or vehicle connected devices (Tran, Tran and Nguyen, 2014) will have real-time information about their load utilization. In current pre-series π-transporters the workload and thus the available space is detected either via RFID/NFC[1] technology (De Wilde, 2004), Bluetooth (Meller, Ward and Gesing, 2020), motion sensors (Knuepfer, 2007) or camera systems (Calver, Cobello and McKenney, 2008). This paper evolved the state of research concept as an artifact that considers the π-nodes as routers in a way that it distributes and replicates real-time data to the π-nodes in order to enable more effective routing decisions. This real-time data is provided by vehicles, or so-called π-transporters, on the road. Therefore, a second artifact will be designed in which π-transporters take over the routing role. In order to be able to take a holistic perspective on the routing topic, the goods that are actually to be moved, the so-called π-containers, are also designed as routing entities in a third artifact. These three artifacts are then compared and evaluated for the consideration of real-time traffic data. This paper proposes π-transporters as routing entities whose software representatives negotiate freight handover points in a cloud-based marketplace. The implementation of such a marketplace also allows the integration of software representatives for stationary π-nodes, which contribute their location and capacity utilization levels to the marketplace. The result makes a valuable contribution to the implementation of the routing component as a part of the Physical Internet framework.

---

[1] RFID/NFC: Radio-Frequency IDentification / Near-Field Communication

## *1. Introduction*

The demand for transport will continue to rise strongly in the coming decades. "*Global freight demand will triple between 2015 and 2050, based on the current demand pathway*", as predicted in the Transport Outlook 2019 (Forum and International Transport Forum, 2019, p. 36). The majority of goods in Germany are transported by trucks. This corresponds to about 72 percent of freight traffic in tonne-kilometres in 2018 (*Güterverkehr 2018*, 2019), of which 37 percent are empty runs (*Verkehr deutscher Lastkraftfahrzeuge*, 2018). Based on the current traffic situation in the existing transport infrastructure, an increase in traffic in the predicted scale is hardly conceivable. Hence, a revolutionary change in transport efficiency is urgently needed. One approach to tackle this change is to transfer the successful model of the Internet (in the following Digital Internet, DI) for data exchange to the physical transport of goods: The so-called Physical Internet (PI, or π). The idea of the Physical Internet is *"a vision of how physical objects might be moved via a set of processes, procedures, systems and mechanisms from an origin point to a desired destination in a manner analogous to how the Internet moves packets of information from a host computer to another host computer"* (Franklin, 2016).

With the PI, methods of a very established information network, the DI, are transferred to physical goods transport. This requires radical changes in current processes, which have an impact on the software and hardware of the involved network elements. As a basis for the following work, PI-specific terms need to be defined.

**Definitions**
The key physical elements, or also entities, in the Physical Internet include π-nodes, π-containers and π-transporters, as introduced by its inventor (Montreuil, 2012). In this paper, the subgroup π-transporter is considered as representative for π-transporter. To use the terms as clearly as possible, these entities are defined below.

Definition 1-1: **π-node** 
A π-node represents a connection point in a network. A π-node is characterized by at least two connections to other network elements, such as other π-nodes. In general, a π-node has the ability to detect, process and forward transmissions for other network nodes.

Definition 1-2: **π-container** 
A π-container encloses freight in such a way that it is made transportable according to its requirements. A π-container also has information about the type of goods, and their source and destination of transport.

Definition 1-3: **π-transporter**
A π-transporter is a moving vessel which enables one or more π-containers to be transported. In most cases these are road-bound vehicles such as cars, vans or trucks. Together with π-conveyors and π-handlers, π-transporters belong to the group of π-movers.

**Where the Problem lies: The challenge of π-nodes as routers**

Why is it relevant to re-think the routing nucleus of the PI? In the Digital Internet, billions of network nodes around the world are interconnected. Messages, consisting of data packets, are not transported along a predetermined route, but only to the nearest node, which then decides to which node it will forward the data packet next (Kaup and Neumayer, 2003). Hence, the nodes in the Digital Internet take over the routing function. Each network node has a forwarding table, which gives it the competence to identify the next best node for forwarding as shown in figure 1. These tables contain the next possible forwarding hubs and the number of steps to the destination, so-called hops.

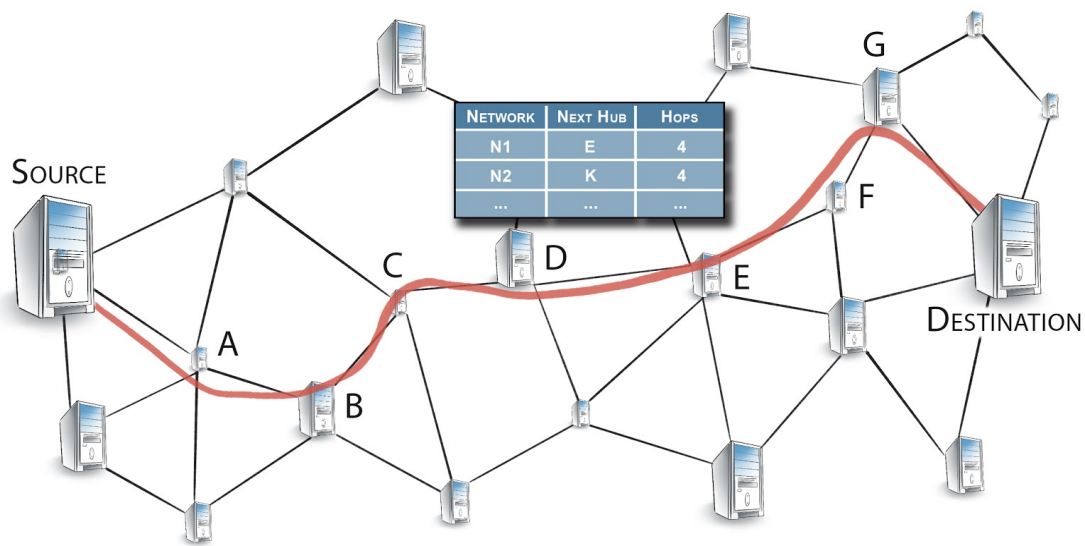

*Figure 1: Routing within the Digital Internet via forwarding tables (own visualization)*

The one-to-one transfer leads to the fact that nodes in the Digital Internet will correspond to π-nodes in the Physical Internet. These π-nodes, or so-called transshipment hubs, have two tasks in this concept: the physical handling of goods and the intelligence to identify the next suitable hub for further transport of the goods. A major weakness of this concept regarding the routing intelligence of the π-nodes is that they do not have any real-time traffic data. According to the current status of the concept, information such as the distance to the next transshipment hub and its accessibility will be used to identify the next step of the transport chain. These criteria are only of limited suitability for successful routing. To improve the routing competence of the hubs, real time traffic bandwidth must be considered. It can be assumed that vehicles or vehicle connected devices will have real-time information about their load utilization in the near future (Tran, Tran and Nguyen, 2014). This leads to the central Research Question (RQ) of this paper:

*Where and how can the routing decision for road-based vehicles be made if real-time data about transport vacancies might be taken into account over the whole PI transport system?* that is solved with the following Sub-Questions (SQ):

- $SQ_1$: How can the concept of π-nodes as the routing elements be extended in order to use real-time data about transport vacancies?

- $SQ_2$: Are there alternatives to π-nodes as routers in order to fulfill the requirement of real-time traffic data usage for routing?

- $SQ_3$: How valuable are these concepts in respect to their routing ability?

The structure of this paper follows the proven process of a design science approach (Wieringa, 2014). Related work and previous research are presented in Section 2. Then, Design Science Research is conducted, consisting of two elementary process steps: 'Build' and 'Evaluate'. Within the Build process, activities are performed that produce design artifacts that are able to use live traffic data as input for the routing within the PI, as described in Section 3. The subsequent 'Evaluate' process in Section 4 evaluates the design artifacts through pre-defined success criteria and provides feedback on it (Österle *et al.*, 2011). Then, Section 5 gives a summary of the results in the form of a concluding statement and provides an outlook on recommended further research on this topic.

## *2. Related Research*

Research related to the Physical Internet dates back to the year 2009, starting with the idea of Montreuil to transform the Digital Internet to a Physical Internet (Montreuil, 2012). Together with Ballot and Meller he wrote a textbook that describes a lot of facets of this transformation (Ballot, Montreuil and Meller, 2014). During the past years and mainly communicated through the proceedings of the International Physical Internet Conference (IPIC) many papers have been published on the idea. They also contain routing mechanisms within the PI. Furthermore, working papers from a European research group, ALICE[2], became available. Based on a systematic literature research regarding the PI, issues related to 'routing mechanisms' and any kind of 'intelligent π-elements', the overview in Table 1 was created.

*Table 1: Results of Systematic Literature Review*

|  | ≤ 2013 | 2014 | 2015 | 2016 | 2017 | 2018 | 2019 | 2020 |
|---|---|---|---|---|---|---|---|---|
| Textbooks | - | - | 1 | - | - | - |  | 1 |
| IPIC Conference Papers | - | 1 | 1 | 1 | 1 | 2 | 2 |  |
| Science Direct | 1 | - | - | 2 | - | - | 3 |  |
| SpringerLink | 2 | 1 | 2 | 2 | - | 2 | 3 | 1 |
| Elsevier | 3 | - | 1 | - | - | - | - | - |
| Emerald Insight | - | - | - | - | 1 | 1 | - | 1 |
| Cornell University | - | - | - | - | - | - | - | 2 |
| Working papers ALICE | - | - | - | - | 1 | - | - | 1 |
| Patents | - | - | - | - | 1 | - | - | - |
| Other publications | 2 | - | - | 1 | 2 | - | 1 | - |
| Total | 8 | 2 | 5 | 6 | 6 | 5 | 9 | ≥ 6 |

---

[2] Alliance for Logistics Innovation through Collaboration in Europe

All these sources describe the routing function as an abstract layer or assume that the hubs, or so-called π-nodes, in the PI will take over this function (Ballot, Gobet and Montreuil, 2012). *'Along the whole transportation process through the logistics network, the loading unit is connected and interacts with the different logistics nodes' (Liesa et al., 2020, p. 37)*. To date, there is little thought about which criteria are suitable to determine the next best π-node. Previous research suggests that criteria such as 'distance to the next transshipment point', 'reachability of the next transshipment point' and the 'probability of further transport' can be used (Montreuil, 2012), but the point of efficient routing needs further research (Sternberg and Norrman, 2017). As an alternative, flow control logic and network management applications could be implemented by a cloud solution (Ballot, Montreuil and Meller, 2014). As recommended in (Becker, 2012), patents were also included in the literature search. One patent (Kaup, 2017b) describes a holistic cloud solution that holds information about all traffic bandwidth and related vacant cargo space and thus gives the freight container and its vehicle a signal when a modality change is considered suitable. Although this takes place locally at π-nodes, it is controlled by freight container representatives within a traffic cloud. Bandwidth information might be collected from π-transporters on the road (Kaup and Demircioglu, 2017). Due to this alternative, scientific databases were searched for the term 'cloud logistics' and the results were included in the 'Build' and 'Evaluation' phase (Ehrenberg and Ludwig, 2014) (Glöckner, Ludwig and Franczyk, 2017) (Ludwig, 2014).

This paper extends the current solution 'π-nodes as routers' and adds other perspectives to the question of where to position routing intelligence. It proposes to use the π-transporters as routers in combination with a cloud-based virtual marketplace.

## 3. Design of Entities for Routing

Design science research is about artifacts in a context (Wieringa, 2014). In the following subsections, each of the elementary entities were put as design artifacts in the context of the routing role. Based on the insights gained in the analysis of existing research, the existing concept of 'π-nodes as routers' will be extended by the integration of real traffic data into the routing decision process. Two further artifacts were as alternatives to the extension of the existing approach. The 'Build' process of the designed artifacts was gained through a joint workshop at 'Mercedes-Benz Innovation Studio' with engineers and researchers in the automotive sector, the field of communication networks and in the world of logistics and telematics.

### 3.1 Extension of the concept π-nodes as routers

The entity π-nodes in the role as routers corresponds most likely to a one-to-one transfer from the routing mechanism of the Digital Internet to the world of physical objects. On the Digital Internet, a distinction is made between static and dynamic routing (Badach and Hoffmann, 2019). In the original context, static routing specifies a defined route definition for data exchange across different nodes. In the physical world, this can be compared with intermodal contract logistics that follow determined ways. This type of routing therefore already seems to be well implemented in the world of physical objects. The other type of routing on the DI is the dynamic routing. Dynamic routing means that the network nodes are responsible for finding the best route for individual data packages. This is done on the basis of criteria such as cost or transmission time. In order to do this, the network nodes must have knowledge of the cost or transmission time of the respective partial routes among themselves. This routing knowledge information is replicated to the network nodes via the so-called Border Gateway

Steffen Kaup, André Ludwig, Bogdan Franczyk

Protocol (BGP) (Badach and Hoffmann, 2019) as described in the introduction. The π-nodes of the Physical Internet correspond to the network nodes in the Digital Internet. In this model, transshipment points, such as rest stops and forwarding agents' yards, decide how the journey might continue for a π-container. The π-transporters (vehicles) would follow instructions set by the π-nodes, transmitted to them e.g. via a fleet management system (Liesa et al., 2020). For this, the π-nodes first need information about which possible routes are available or which possible π-transporters still have free capacity on these routes. This information is collected from the π-transporters.

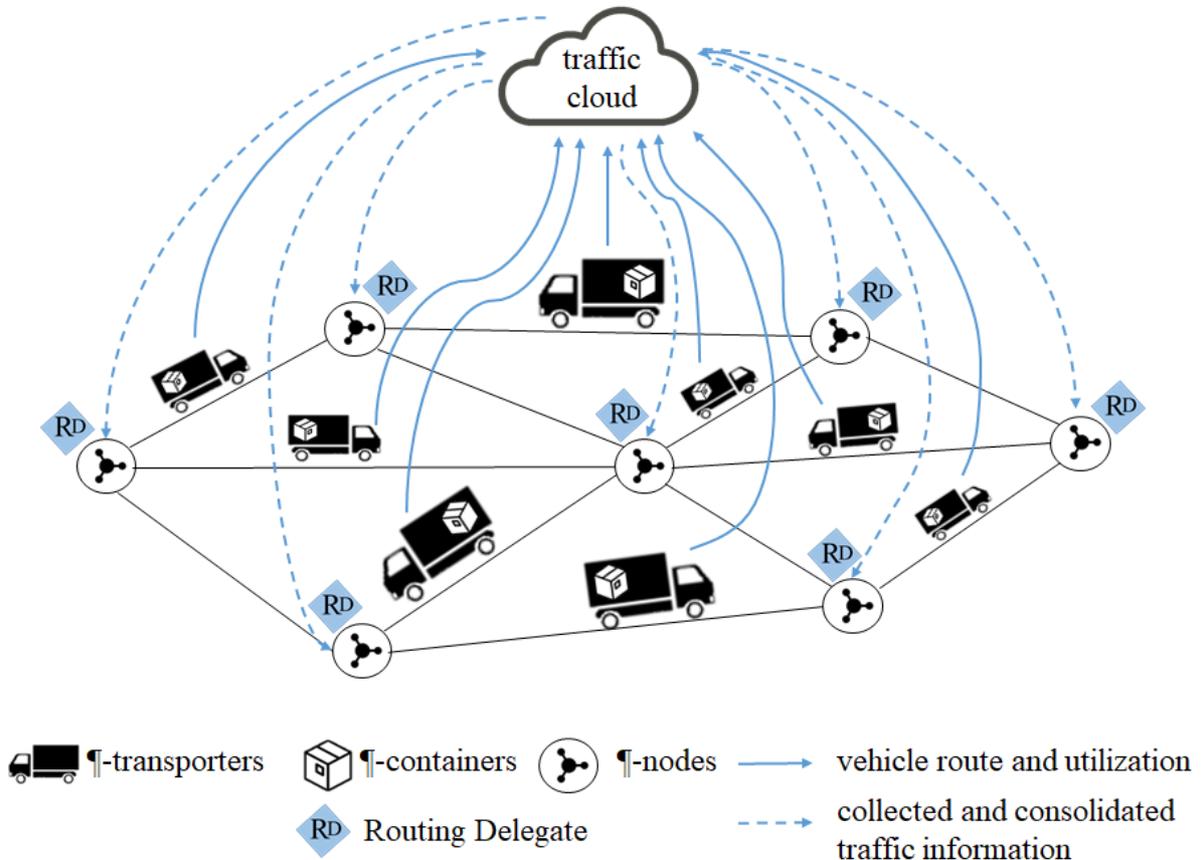

Figure 2: System of π-nodes empowered through traffic relevant data

They would either have to transmit this data to all surrounding π-nodes or send it to a common traffic cloud, which replicates the information to the π-nodes in analogy to the BGP. The routing decision is then finally made by the π-nodes. The π-nodes can be seen as Routing Delegates ($R_D$) of the traffic cloud (Gamma, 1995). Figure 2 visualizes an example of a cloud-supported system with 'π-node delegates' as routers.

### 3.2 Alternative Entity π-transporters as routers

With π-transporters as routers, the vehicles on the road are in the role of decentralized real-time decision making. They have knowledge about their planned routes and ideally about their load conditions. Tracking of the load status of a vehicle can be realized via a small onboard network, e.g. RFID/NFC (De Wilde, 2004), Bluetooth (Meller, Ward and Gesing, 2020), motion sensors (Knuepfer, 2007) or camera systems (Calver, Cobello and McKenney, 2008). This is visualized in Figure 3 by displaying a small network symbol with three arcs in the vehicles. Also, transporters are marked as Routing-Hubs ($R_H$) in this Figure. Through connecting to each other via car-to-x communication, or so-called mesh networks (Jiang *et al.*, 2020), π-transporters are able to exchange relevant information with each other in order to negotiate freight exchange points among themselves. Such a vehicle network can be seen as a kind of 'trading venue' where next appropriate routes and necessary modality switches can be negotiated. This network might be implemented in a distributed way among the vehicles within the vehicles mesh network, e.g. using Distributed Ledger Technologies (DLT), such as blockchain (Mollah *et al.*, 2020).

This network would also be able to heal freight transport routes in case one or more of the vehicles would fail or are delayed in a traffic jam. Information on the location of hubs is not prone to change as frequently as that of traffic, so that it can be integrated into existing map material with acceptable effort. Through this technology, dynamic transfer points outside the range of regular hubs might also be negotiated among the vehicles within the mesh network.

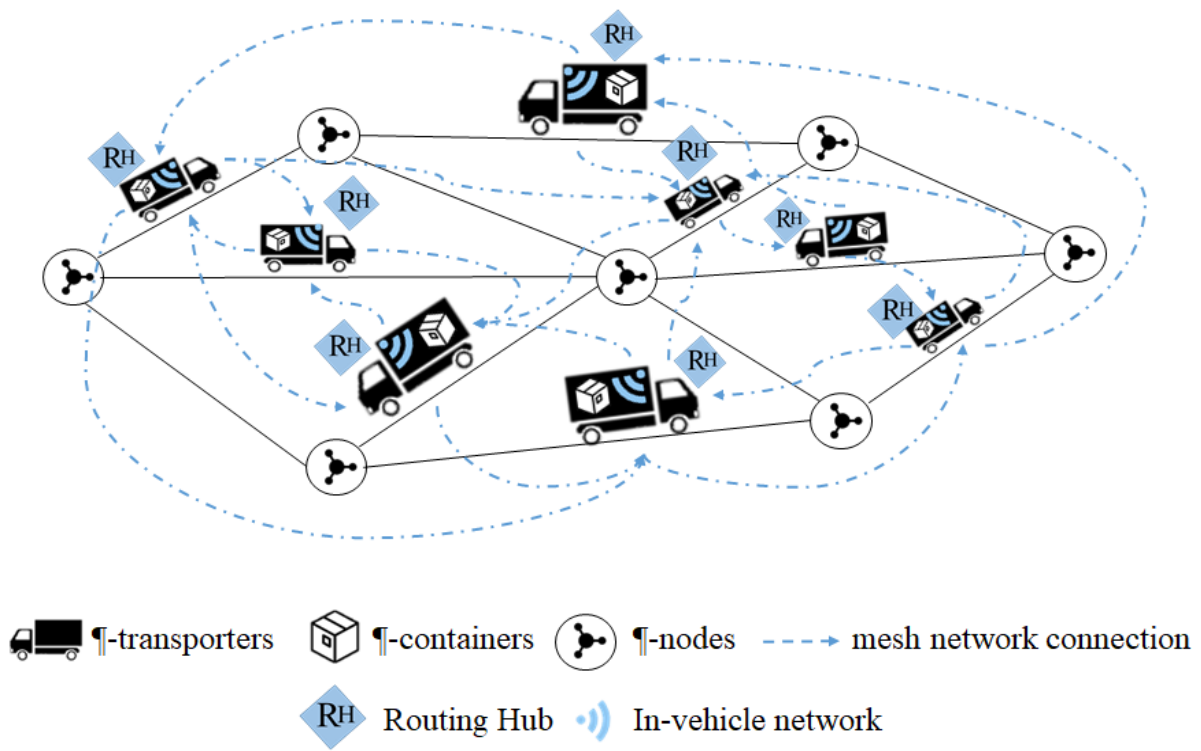

*Figure 3: Example of a mesh network of π-transporters in the role of routing hubs*

Hence, these transfer points do not necessarily have to correspond to static transshipment yards, but also to mobile hubs or dynamic rendezvous points. This makes the system highly flexible and efficient. In a final expansion stage, freight could already be exchanged during the ride, similar to some approaches under study for passenger transport. This routing artifact is easily scalable, so that more and more π-transporters might be enabled as mobile hubs. These vehicles trade among themselves for more or less capacity and thus earn additional money during operation. It can be said that this will enable tasks to be taken over by the π-transporters that were previously performed by freight exchange companies.

### 3.3 Alternative Entity π-containers as routers

To have π-containers as the routing entity would mean that the containers itself could make routing decisions. However, for this they need information about traffic conditions and unused capacities within π-transporters. Even if they could negotiate directly with vehicles, quasi like a hitchhiker, the onward journey would not be secured. For this reason, every π-container has a software agent that represents himself in a common traffic cloud, like a digital twin (Hofmann and Branding, 2019). Hence, the traffic cloud, also called the Routing Brain ($R_B$), orchestrates the utilization of the π-containers in a holistic way (Kaup, 2017b), as visualized in Figure 4.

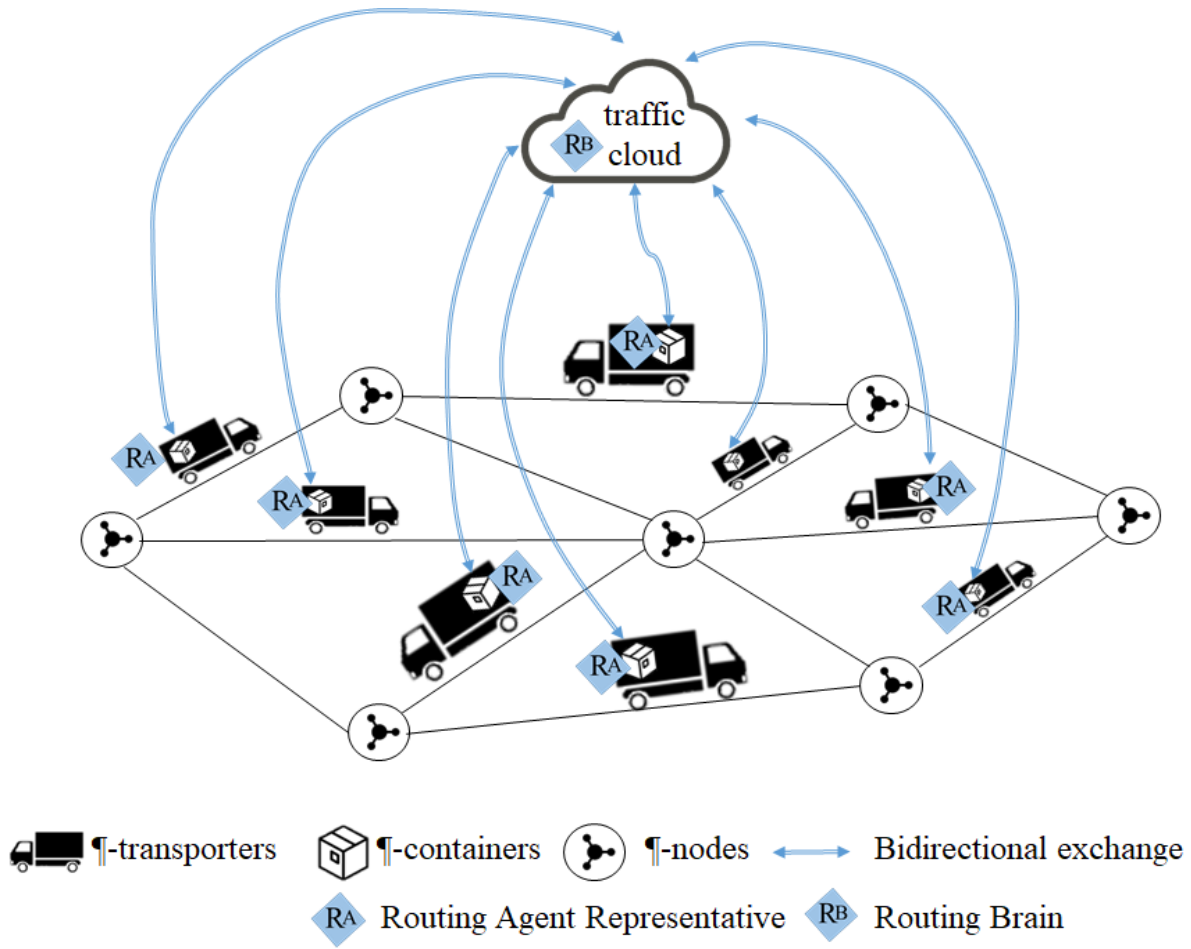

*Figure 4: System of π-containers and their Routing Agents*

A bidirectional exchange between π-containers and the traffic cloud is necessary in order to inform the traffic cloud about slow-moving traffic or possible accidents, which requires a network with high bandwidth and ultra-low latency (5G). Possible lacks or dysfunctionality of traffic could be identified on feedback from π-containers to the traffic cloud concerning their current location information.

This approach corresponds to the mobility behavior of people who book common modes of transport or mobility via an app (e.g. to reserve a seat on a train or a taxi on call). By using a platform, e.g. Moovel[3], transport requirements are entered and then suitable intermodal routes are calculated by the cloud platform depending on time and cost. In a similar logic, mobile π-containers could be routed and operated via apps (Tran-Dang and Kim, 2018). Since containers themselves cannot enter data into an app, they need a software agent as a representative in the cloud ($R_A$, see also Figure 4) (Zhou and Lou, 2012). Such a container representative communicates with the π-container and steers it through the traffic network

---
[3] Moovel Group GmbH (new name: REACH NOW), Software Company, Stuttgart, Germany

using the cloud information as the Routing Brain. For this purpose, each π-container must be equipped with a transmitting and receiving unit. Clear standards are required to integrate as many π-containers as possible in this system. The high-quality and expensive π-containers must also be reused. If possible, no empty π-container should go back anywhere. In the case of mixed operation, i.e. if existing dedicated traffic should also be used, the cloud must also have information on vehicle utilization levels, similar to π-nodes as routers.

In summary it can be said that all artifacts make routing decisions based on real-time data about transport vacancies. This kind of data is collected and made available by π-transporters in all solutions. The differences of the artifacts lie in which π-element the routing decision is made and whether this is done peer-to-peer or by software-representatives in a cloud solution.

## 4. Routing Entity Evaluation

In this Section, the 'Evaluate' process takes part, that means the designed three artifacts have been assessed in their routing context. First, evaluation criteria were determined. The criteria make it possible to assess the robustness and rigor of the designed artifacts as well as the efficiency gain and accessibility to other providers. Then, the artifacts were validated against these criteria by an empirical study.

### 4.1 Criteria and Methodology for Evaluation

The objective of evaluation methodology is to make sure that satisfactory progress is being made towards fulfilling deliverables and reaching relevant contributions to the PI. In literature, evaluation *'serves the purpose of deciding whether or not to acquire or develop a technology, or the purpose of deciding which of several competing technologies should be acquired or adopted (Venable, Pries-Heje and Baskerville, 2016, pp. 77–89)'*. The method in process used in this paper is the qualitative analysis of expert interviews. In order to be able to compare the artifacts with one another as well as possible, the authors chose the deductive category application (Mayring, 2014). In order to perform this and to ensure the rigor of the research, accepted criteria are needed to assess how well the constructed artifacts fit into the routing context (Österle *et al.*, 2011). But how to find accepted criteria? The reason for doing research on the PI is the respected outcome of a higher degree of 'Efficiency' which qualifies this criterion for the catalogue In addition, the research group ALICE worked out 'Seamlessness' and 'Scalability' as important requirements for logistics networks (Liesa et al., 2020). A seamless transition from one mode of transport to another is necessary for freight that requires special handling, e.g. refrigerated food or medicine. Scalability indicates how easy the transport network can be extended, e.g. by new routing elements. Discussions with experts from freight forwarding companies lead to the finding that scalability is not sufficient. They require interoperability of different players cross-brand. As an example, the technology company Apple makes it easy for users to add new Apple components within its own product environment, but as hard as possible to add components from competitors. To enable a great possible effect of the PI, we need the interoperability of different players and companies. For this, the criteria 'Openness' was added. Discussions with customers led to the finding that the newly designed system must be at least as good as current-world logistics. To ensure this, the criteria 'Costs', 'Time' and 'Reliability' were added, as shown in Table 2.

*Table 2: Criteria for Routing Entity Evaluation*

| Criterion | Description |
| --- | --- |
| Scalability | Scalability indicates how easy the transport network can be extended, e.g. by new routing elements. |
| Openness | Ensures cross-brand accessibility of π-elements and the interoperability of different players, such as large shipping companies, smaller vehicle fleets, solo entrepreneurs. |
| Efficiency | Indicates how efficiently transport and routing take place. This is done by estimating the average capacity utilization rate of the modes of transport. |
| Costs | The cost of implementing a functional self-routing system consisting of the necessary adaptations to the relevant π-elements, such as π-nodes, π-transporters or the cost of developing a necessary cloud platform. |
| Time | Estimated qualitative transport time of representative end-to-end connections of a standardised container within the network. |
| Reliability | Reliability of the onward transport of goods from one hub to the next hub. |
| Seamlessness | The retail customer is not affected by freight changes in modes and routes due to the comprehensive and fully interconnected network. In addition, cold chains can also be ensured by reducing the probability of goods being detained during cargo handling. |

In order to achieve the greatest possible gain in knowledge, a qualitative empirical approach with experts was chosen. For this, the designs of the three artifacts were discussed with experts, who imagine how such an artifact will interact with the problem context of routing freight within a Physical Internet. Then, they predicted what effects regarding the determined criteria they think this would have.

### *4.2 Results of Evaluation*

Evaluation by expert opinion only works if the experts understand the artifacts, imagine realistic problem contexts, and make reliable predictions about the artifacts in context. Hence, it was not a trivial task to find the right experts to evaluate the designed artifacts. The requirements for the interview partners were that they had to have both expert knowledge in functioning of the DI as well as in transport logistics. The central statements of these experts (N=9) are shown in a distilled form in Table 3. The interviewed experts were divided into the categories 'professors or chairs of renowned universities'[U], 'founder of highly innovative startups or CEO's of consulting companies'[C] and 'research leaders within the automotive industry'[A]. The superscripts (U, C, A) on the central statements in the evaluation matrix indicate the category to which the expert who made this statement belongs.

*Table 3: Evaluation Matrix of Routing Entity Artifacts*

| Criterion | Routing Entity #1: π-nodes | Routing Entity #2: π-transporters | Routing Entity #3: π-containers |
|---|---|---|---|
| Scalability | + As soon as a traffic cloud is implemented, more and more π-nodes can be added[U]<br>- π-nodes still must be provided with relevant traffic information[U] | + Most vehicles already have communication technology on board[A]<br>- Limited suitability for the long haul because of limited communication range to other vehicles[C] | + Once an infrastructure is agreed, π-containers can be added easily[U]<br>- A global communication standard is needed for all containers[C]<br>- 5G is required[U] |
| Openness | + Other modalities are convenient to include, because stationary hubs are often located at railway stations or ports[C]<br>- Most of π-nodes are privately owned, agreements of use are difficult to conclude[A] | + New routing elements in form of π-transporters can be easily integrated[A]<br>- Building Ad-hoc networks with other modalities, like trains or ships is seen as a challenge[C] | + Good, under the condition that the cloud-platform is open and barrier-free for all kinds of π-container representatives[U]<br>- In a mixed operation with existing traffic, an additional connection from π-transporters to the cloud is required[A] |
| Efficiency | - Only static π-nodes can be used[U] | + Dynamic π-nodes possible (rendezvous points)[A] | + Dynamic π-nodes possible (rendezvous points)[A] |
| Costs | - Collecting traffic information and replicating them to π-nodes is complex[U] | + Many π-transporters already have telecommunication systems, it is "just" a software topic[A] | - Expensive, as each π-container would have to be equipped with a long-range telematic unit[C] |
| Time | o Can only be evaluated after implementation or simulation[U,C,A] | | |
| Reliability | + The responsibility clearly lies with the π-nodes. Hence, reliability is seen as high.[C] | o Reliability depending on defined communication and negotiation standards[A] | o Reliability depending on Software Agents in the Cloud and the integration of existing traffic[A] |
| Seamlessness | - The π-nodes must get information about the transport network from somewhere[U] | + Good, if manufacturer-independent standard exists and DLT platform works[A] | + Very good, if non-proprietary container-standard exists[C] |

The findings of the designed artifacts indicate that each of the solutions has advantages and disadvantages. The most technically mature solution is not always the one that can be quickly and safely established on the market. In the following subsections the obstacles and opportunities of the routing artifacts are discussed.

*Opportunities and obstacles of 'π-nodes as routers'*

All the interviewed experts agree that it is necessary to include current information on traffic volume and vehicle utilization in the routing process. The artifact 'π-nodes as routers', as extended in Section 3.1, rises and falls with the ability to replicate traffic and vehicle related information on the π-nodes. Also, this entity design is limited to stationary hubs as routing elements. This may seem simple at a first glance, but it should be noted that most of the π-nodes are privately owned and operated, because almost all logistic hubs are dedicated to a freight forwarding company, e.g. DHL. It must be ensured that these possible π-nodes can also be used without manufacturer discrimination, perhaps even with political support. With π-transporters or π-containers as routers, dynamic transfer points are theoretically possible. But, how does the goods turnover look like there? Since there are no stationary handling robots on site of π-nodes, the physical handling of goods either has to be carried out by humans or the π-containers have to be mobilized in some way, for example through mobile delivery robots (Canoso, Binney and Rockey, 2017).

*Opportunities and obstacles of 'π-transporters as routers'*

Six of the nine experts surveyed, particularly those from the automotive industry, are convinced that the concept of smart vehicles, as introduced in Section 3.2, would ease the issue of protocol development and complexity quite considerably. In the artifact 'π-transporters as routers', π-transporters (vehicles) negotiate the next best mode of transport for containing freight among themselves, just as autonomous vehicles will have to negotiate the right of way with each other in the future. It has the charm that (almost) every vehicle (as the most common implementation of π-transporters) has a connection to the Internet and therefore to another vehicle. If the vehicle itself should not have this kind of connection, there is often a driver sitting in it, who has a connection to the Internet via personal devices. Hence, π-transporters could be connected to each other easily, e.g. via an app as described in (Tran, Tran and Nguyen, 2014). In order to negotiate the onward transport of goods between vehicles, they must have information about the π-containers and their destinations. If the π-transporters are intelligent, then an entirely different concept for control can be utilized that really simplifies the replication problem, resulting from the concept of π-nodes as routers. They still need to understand the state of the network, but rerouting and load management could be organized by the vehicles themselves, e.g. by using a blockchain-backed broker platform (Meyer, Kuhn and Hartmann, 2019). Tracking and tracing of π-containers within the vehicles can be realized through technologies like RFID (De Wilde, 2004) or motion sensors (Knuepfer, 2007). Hence, the π-transporters act as a 'kind of mobile hub' with acceptable efforts and costs. By using exchangeable body capsules, the π-transporters entity concept would also be transferable to passenger transportation and mobility solutions (Froböse, 2012).

*Opportunities and obstacles of 'π-containers as routers'*

The concept of 'π-containers as routers' would be as if packets on the Internet were made intelligent by their cloud representatives and could, therefore, dynamically manage their movements through the network. In the eyes of many experts, this concept requires a complex sending and receiving unit at the side of the π-containers, which must be able to connect to their cloud representatives. Likewise, the network development for 5G must be sufficiently progressed. This requires much effort and leads to high costs. In a mixed operation, the cloud must first be supplied with holistic traffic data, as described in Section 3.1. The routing algorithm must also ensure that the system does not return these expensive π-containers empty again. As an alternative approach, π-containers could also become an integral part of mesh-networks, like mentioned in (Ballot, Montreuil and Meller, 2014) as follows: '*A container that becomes an integral part of the Internet of Things, along with its handling and storage equipment and transportation, then allows these to be coordinated through machine-to-machine communication.*' But, it will take a long time to develop a smart container network architecture across different manufacturers of π-containers (Marino *et al.*, 2019) or to connect π-containers directly to π-transporters or π-hubs with ultra-low latency via 5G (Pagano et al., 2019).

According to the highest number of advantages and the lowest number of disadvantages, the artifact 'π-transporters as routers' is considered the most appropriate solution. Hence, this paper proposes to use π-transporters, in particular vehicles on the road, as routers. Either the vehicles communicate with each other via ad-hoc mesh networks or software representatives perform this task for them in a kind of logistics cloud (Glöckner, Ludwig and Franczyk, 2017). This depends on how much computing power is available in the vehicles and whether they can communicate with each other across manufacturers. Artifact 'π-transporters as routers' and artifact 'π-containers as routers' could be combined to overcome the weakness of the low range of the vehicle mesh network. Five of the experts consider it useful to have all traffic relevant data in one place. Hence, a cloud solution is preferred, which serves as a virtual marketplace for freight exchange between π-transporters. Most of the transshipment operations will continue to take place at stationary hubs. The implementation of a virtual marketplace also allows the integration of software representatives for stationary π-nodes, which contribute their location and capacity utilization levels to the marketplace. The opening of the marketplace for π-container representatives can also be considered, so that customers can directly contribute freight with transport needs. To better predict routes and thus improve long-distance routing, Artificial Intelligence in combination with Game Theory methods could further increase the effectiveness of the virtual marketplace. This could further optimize the multi-agent system in order to support software agents finding their best negotiation partners for taking over the freight for onward transportation.

## 5. Conclusions and further work

In the logistics of physical objects, a big challenge lies in non-use of free transport capacities. A method, similar to dynamic routing on the Digital Internet, can be expected to increase efficiency in the whole transport chain. Current research identified and determined the hubs in charge of routing freight efficiently through a Physical Internet. This paper addresses the as yet unsolved problem of how to identify the next best appropriate hub for onward transport based on real-time traffic data. Hence, this paper answers the central Research Question (RQ) of where and how the routing decision for road-based vehicles can be made if real-time data about transport vacancies might be taken into account for the routing schemes within the PI. To solve this problem, the existing concept of routing π-nodes was extended with the supply of real-time traffic data by collecting it from π-transporters and storing them into a cloud. This cloud provides and replicates the data about transport vacancies to the π-nodes and therefore answers $SQ_1$. This led to the design of an alternative artifact that sets the π-transporters directly in the routing role. In order to negotiate transfer points with other transporters, a mesh network between them was designed. Last but not least, the containers themselves were put into the routing role via software-representatives that negotiates further transport possibilities in something like a cloud-based marketplace. These two designed alternatives, π-transporters and π-containers as routers, answer $SQ_2$. Then, the three artifacts

Steffen Kaup, André Ludwig, Bogdan Franczyk

were evaluated due to their routing competence regarding criteria, derived from previous literature and research, that address $SQ_3$. As a result, there is no solution that is the most suitable in all aspects, but the evaluation proposes π-transporters as routers. The reasons for this are twofold: the π-transporters collect the traffic data due to identify available transport bandwidth and already have suitable telematics devices on board in order to negotiate possible transfer points for freight. A key finding of this paper is that real-time data about available vacancies in π-transporters, collected by a swarm of π-transporters, is key as a reasonable basis for routing freight in an effective way through the PI. Either the vehicles communicate with each other via ad-hoc mesh networks or software representatives perform this task for them in a common logistics cloud. This depends on how much computing power is available in the π-transporters and whether they can communicate with each other across manufacturers. From a programmer's perspective, it is of decisive advantage to have all data in one place. Therefore, a cloud solution is preferred, which serves as a virtual marketplace for freight exchange between π-transporters. In the same way, this marketplace could be extended by software-representatives of π-nodes and π-containers and thus let them participate in the negotiation process of the intermodal transport chain.